\documentclass[a4,aps,twocolumn,showpacs,preprintnumbers,amsmath,amssymb]{revtex4} 
\usepackage{dcolumn}
\usepackage{graphicx}
\usepackage{amsfonts} 
\usepackage{epstopdf}
\usepackage{color}
\usepackage{here}

\begin{document} 
\title{Vortices in a toroidal Bose-Einstein condensate with a rotating weak link}
\date{\today}
\author{A.I. Yakimenko$^{1}$, Y.M. Bidasyuk$^{2,3}$, M. Weyrauch$^{2}$, Y.I. Kuriatnikov$^1$, S.I. Vilchinskii$^1$}
\affiliation{$^1$Department of Physics, Taras Shevchenko National University of Kyiv, Volodymyrska Str. 64/13, Kyiv 01601, Ukraine \\
$^2$Physikalisch-Technische Bundesanstalt, Bundesallee 100, D-38116 Braunschweig, Germany \\
$^3$ Bogoliubov Institute for Theoretical Physics, National Academy of Sciences of Ukraine, Kyiv 03680, Ukraine}
\begin{abstract}
Motivated by a recent experiment 
[K.C. Wright et. al. Phys. Rev. Lett. 110, 025302 (2013)], we investigate deterministic discontinuous jumps
between quantized circulation states in a toroidally trapped Bose-Einstein condensate. These phase slips are induced by vortex excitations created by a rotating weak link. We analyze influence of a localized condensate density depletion and atomic superflows, governed by the rotating barrier, on the energetic and dynamical stability of the vortices in the ring-shaped condensate.
We simulate in a three-dimensional dissipative mean field model  the dynamics of the condensate using parameters similar to the experimental conditions.
Moreover, we consider the dynamics of the
stirred condensate far beyond the experimentally explored region and reveal surprising manifestations of complex vortex dynamics.


\end{abstract}

 \maketitle

\section{Introduction}


Experimental realization of ultracold atomic ``circuits'' with a  tunable weak link ~\cite{PhysRevLett.99.260401,PhysRevLett.106.130401,PhysRevLett.110.025302,PhysRevLett.113.045305} opens an intriguing perspective for fundamental studies of the superfluidity at the new level of control, manipulation, and measurement.
A super-fluid ring with a weak link 
as well as a super-conducting circuit with a Josephson junction
can act as a nonlinear interferometer,
allowing the construction of high-precision detectors
such as superconducting quantum interference
device (SQUID) magnetometers and superfluid gyroscopes.
The fast response of the condensate phase
winding on the variations in the rate of barrier angular velocity makes the atom SQUID ideal for future highly sensitive
rotation-measurement devices \cite{PhysRevLett.111.205301}.




Bose-Einstein condensates (BECs) in toroidal traps are the subject of many experimental and theoretical investigations
\cite{PhysRevA.66.053606,BenakliEuL99,Brand01,Das2012, PhysRevA.64.063602,PhysRevA.74.061601}
which study persistent currents \cite{PhysRevLett.99.260401,PhysRevLett.110.025301,PRA2013R}, weak links \cite{PhysRevLett.106.130401,PhysRevLett.110.025302}, solitary waves \cite{Brand01,Berloff09}, and the decay of the persistent current via phase slips \cite{PhysRevA.86.013629, PhysRevA.80.021601,Piazza2013} in this doubly connected topology. A persistent atomic flow in a toroidal trap can be created  by transferring  angular momentum from optical fields \cite{PhysRevLett.99.260401,PhysRevLett.110.025302} or by stirring with a rotating barrier~\cite{PhysRevLett.110.025302,Wright2013}.


A controllable creation of the phase slips was achieved in experiment recently ~\cite{PhysRevLett.110.025302} by
increasing the rotation rate of the barrier from zero up to $\Omega / 2\pi \le 3$ Hz.
The experimental observations were treated in Ref. \cite{PhysRevLett.110.025302} within a 1D model
which provides a good insight into the origin of the critical current inside the weak link, even though the width of the annulus in the experiment is of the same order as the radius of the trap. However, the analysis of the excitation and further dynamics of  the vortices, which appear in the critical regime, is beyond the scope of 1D model.  This calls for an analysis  in a 2D or 3D  model.

In the present work we investigate formation of the persistent current via the deterministic phase slips driven by the wide rotating barrier.
The rotating barrier modifies the condensate density, creating a weak link, and induces the superflows in the ring-shaped condensate.
We analyze an influence of both these factors on the energetic and dynamical stability of the vortices in the ring-shaped condensate.
We numerically simulate dynamic of the phase slips in the framework of time-dependent dissipative 3D Gross-Pitaevskii equation (GPE) for the parameters matched to the experimental conditions. We determine the minimal angular frequency and the minimal height of the barrier, which are needed for excitation of the persistent current under conditions of the experiment \cite{PhysRevLett.110.025302}. Furthermore, we consider the dynamics of the stirred condensate far beyond the experimentally explored region and reveal surprising manifestations of complex vortex dynamics.

\section{Model}\label{model}
In the mean field approximation, the dynamics of a system of weakly interacting degenerate atoms  close to
thermodynamic equilibrium and subject to weak dissipation is described
by the GPE \cite{Pitaevskii59,Choi98}
\begin{equation}\label{GPE}
(i-\gamma) \hbar \frac{\partial \tilde\Psi(\textbf{r},t)}{\partial t} = \left[\hat H + \tilde g |\tilde\Psi(\textbf{r},t)|^2 -\mu\right]\tilde\Psi(\textbf{r},t).
\end{equation}
Here $\gamma\ll 1$ is a phenomenological dissipation parameter, and  $\hat H=-\frac{\hbar^2}{2M} \Delta + V(\textbf{r},t)$ the Hamiltonian;  $\Delta$
is the Laplace operator, $\tilde g = 4 \pi \hbar^2 a_s/M$ is a coupling strength, $M$ is the mass of the $^{23}$Na atom, and $a_s=2.75\,$nm is the $s$-wave scattering length. We assume that the temperature of the condensate is well below the condensation temperature: $T\ll T_c$. The chemical potential $\mu(t)$ of the equilibrium state in our dynamical simulations was adjusted at each time step so that the number of condensed particles slowly decays with time: $N(t)=N(0)e^{-t/t_0}$, where $t_0 = 10\,$s corresponds to the $1/e$ lifetime of the BEC reported in Ref.~\cite{PhysRevLett.110.025301}.

The external potential $V(\textrm{\textbf{r}},t)=V_t(r,z)+V_b(\textrm{\textbf{r}},t)$ involves the axially-symmetric, time-independent toroidal trap and the time-dependent potential of a rotating repulsive barrier $V_b(\textrm{\textbf{r}},t)$.

As in Ref.~\cite{PhysRevLett.110.025302} the trap is approximated by a harmonic potential
\begin{equation}\label{toroidal_trap}
V_{\textrm{t}}(r,z)=\frac12 M\omega_r^2(r-R)^2+\frac12 M\omega_z^2 z^2.
\end{equation}
with  $r=\sqrt{x^2+y^2}$.
The trap frequencies $(\omega_r,~\omega_z)=2\pi\times (123,~600)$ Hz correspond to the oscillator lengths $l_r=\sqrt{\hbar/(M\omega_r)}= 1.84\,\mu$m and $l_z= 0.83\,\mu$m.   Number of atoms $N = 6\cdot 10^5$ and chemical potential $\mu/h=2\,$kHz. The radius of the ring trap is $R = 19.23\,\mu$m.

In terms of harmonic oscillator units [$t\to \omega_r t$, $(x,y,z)\to (x/l_r,y/l_r,z/l_r)$, $V\to V/(\hbar\omega_r)$, $\mu\to \mu/(\hbar\omega_r)$] Eq.~(\ref{GPE}) can be written in dimensionless form
\begin{equation}\label{GPE_dimensionless}
(i-\gamma)\frac{\partial \psi}{\partial t} = \left[-\frac{1}{2} \Delta + V(\textrm{\textbf{r}},t) +  g|\psi|^2 -\mu\right]\psi,
\end{equation}
with $\psi = l_r^{-3/2}  \tilde\Psi$ the dimensionless wave-function, $g=4\pi     a_s/l_r=1.88\times 10^{-2}$ the interaction constant. The parameters of the static potential
$V_t(r,z)=(r-R)^2/2+\kappa^2z^2/2$  are the dimensionless radius
$R=10.4$ and the ratio of oscillator frequencies $\kappa=\omega_z/\omega_r=4.88$.


Experimentally~\cite{PhysRevLett.110.025302} an effective time-averaged barrier was created by scanning the radial position
of the blue-detuned laser beam across the condensate, with a
scan amplitude greater than the width of the annulus.
For simplicity, we assume a weak link described by a potential, which is homogeneous in radial direction across the toroidal condensate,
\begin{equation}\label{straigthBeam}
V_b(\textbf{r}_\perp,t)=U(t)\Theta(\textbf{r}_\perp\cdot \textbf{n})e^{-\frac{1}{2c^2}\left[\textbf{r}_\perp\times \textbf{n}\right]^2},
\end{equation}
where  $\textbf{r}_\perp=\left\{x,y\right\}$ is the radius-vector in $(x,y)$ plane; the unit vector $\textbf{n}(t)=\left\{\cos(\Omega t),\sin(\Omega t)\right\}$  points along the azimuth of the barrier maximum. The dimensionless parameter $c=1.85$ was chosen to reproduce a $8\,\mu$m diameter (FWHM) of the  scanning laser beam. The Heaviside theta function $\Theta$ in Eq.~(\ref{straigthBeam}) assures a semi-infinite radial barrier potential starting at the trap axis. Note that the effective width of the barrier is much greater than the healing length at the peak density: $\tilde\xi=\sqrt{1/(8\pi a_s  n_0)}=0.32\,\mu$m, which gives in dimensionless units $\xi=\tilde\xi/l_r=0.175$.

\begin{figure}[ht]
  \includegraphics[width=3.4in]{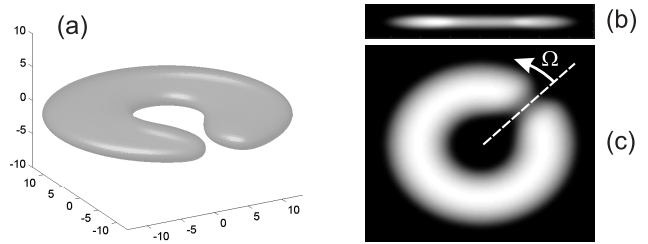}
  \caption{(a) Isosurface of the toroidal condensate density with a  weak link of $U_b/h=2.3$ kHz height. The integrated condensate density  (lighter shades indicate larger column
densities) for $U_b/h=1.3$ kHz: (b) $n(x,z)$ -- side view (size of the image is $40\times 6$ in units of $l_r$); (c) $n(x,y)$ -- top view  (the size of the image is $40\times 40$ in units of $l_r$). The azimuth of lowest density, indicated by the dashed line.}
  \label{WeakLinkScheme}
  \end{figure}

Following the experimental setup~\cite{PhysRevLett.110.025302} the amplitude $U(t)$ of the weak link  varies in time:
During the first 0.5\,s  the height $U(t)$ of the potential barrier ramps up to a maximum value $U_b$. For the next 0.5\,s the amplitude of the barrier remains constant, and during the last 0.5\,s the potential barrier ramps down. During the stirring period the penetrable barrier moves anti-clockwise with constant angular velocity $\Omega$.
Fig.~\ref{WeakLinkScheme}(a) shows an isosurface of the density distribution of the toroidal condensate with the weak link. In Fig.~\ref{WeakLinkScheme}(b) and (c) the y-dependence or z-dependence is integrated out, respectively.

\section{Ground state of the toroidal condensate and energetic stability of the vortex excitations}

In line with the experimental conditions we prepare a nonrotating state in the toroidal trap. We use the imaginary time-propagation method to find numerically the stationary solution of the Eq.~(\ref{GPE_dimensionless}) without dissipation ($\gamma=0$) and without weak link ($V_{b}=0$). The stationary state contains $N = 6\cdot 10^5$ atoms from which we calculated the dimensionless chemical potential $\mu=16.55$.



Quantized circulation in a ring
corresponds to an $m$-charged vortex line pinned to the center of the ring, where the vortex energy has a local
minimum. The true ground state of the
system in a non-rotating trap has charge $m = 0$, however, since the vortex core is confined by a potential barrier to the central hole of the condensate, even multi-charged ($L_z/N=m>1$) metastable vortex states has a very long life-time~\cite{PhysRevA.86.013629}.

It is instructive to estimate the energy variation of a vortex line when it is shifted from the central hole of the toroidal condensate through the weak link. Let us calculate the nucleation energy of a vortex line  as a function of its spatial location, accounting for the
condensate inhomogeneity caused by the trapping potential, the  localized region
of reduced superfluid density caused by the  weak link, and the hole around the vortex core.
Previously, a similar approximation was used for the analysis of the energetic stability of the vortex lines and vortex rings in a single-connected trap \cite{Jackson99,PRA13}. First, using the
imaginary time propagation method, we find numerically the non-rotating
ground state density $n_{\textrm{GS}}=|\Psi_{\textrm{GS}}(\textbf{r})|^2$ of a BEC for the given trapping potential and barrier height $U(t)=U_b=$const,  $\textbf{n}=\{1,0\}$ is the unit vector along the azimuth of the barrier maximum.
Then we imprint the offset vortex line into the ground
state and obtain the condensate density with imprinted vortex line:
\begin{equation}\label{anzats}
n(x,y,z)=A\left[\tanh\left\{\rho/\xi\right\}\right]^{2m}n_{\textrm{GS}}(x,y,z),
\end{equation}
where $A$ is the normalization constant introduced to
preserve the number of atoms, $m$ is the topological charge of the imprinted vortex line. We assume that the vortex line is parallel to the $z$-axis, and $\rho(x,y)=\sqrt{(x-x_0)^2+y^2}$ is the distance from the axis of the vortex core. We estimate the width of the vortex core by the healing length $\xi$ at the peak density of the condensate. The ansatz (\ref{anzats}) interpolates the condensate density properly both in the vicinity of the core, $n\sim \rho^{2m}$ if $\rho\to 0$, as well away from the core, $n\sim n_{\textrm{GS}}$ if $\rho\gg \xi$.
 \begin{figure}[t]
  \includegraphics[width=3.4in]{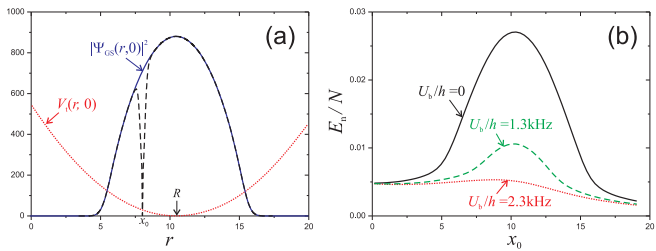}
  \caption{(Color online) (a) Condensate density $|\Psi(r,z=0)|^2$ in the ground state (blue curve), the density distribution with imprinted offset vortex line (shown by dashed curve), and the toroidal trapping potential $10\times V_t(r,z=0)$ (given by dotted red curve). (b) The nucleation energy per particle $E_\textrm{n}/N$ of the offset vortex line shifted at $x_0$ from the axis of the trap in the direction of the weak link lowest density. Shown are the curves for three different values of the barrier height: $U_\textrm{b}=0$ (solid black curve), $U_\textrm{b}/h=1.3\,$kHz (dashed green curve), and $U_\textrm{b}/h=2.3\,$kHz (dotted red curve).}
  \label{NucleationEnergy}
  \end{figure}
The total condensate energy without dissipation is given by the following functional:
$$
E=\int \left\{\frac12|\nabla\psi|^2+V_{\textrm{ext}}(\textbf{r})|\psi|^2+\frac12 g|\psi|^4\right\}d^3\textbf{r}.
$$
Since the nucleation energy, $E_\textrm{n} = E_\textrm{v} - E_{\textrm{GS}}$, is
difference of the energy of the state with imprinted vortex line and  the
ground state energy, the main contribution to the nucleation $E_\textrm{n}$ energy
is given by the kinetic energy of the vortex-induced superflow:
$$E_\textrm{n}\approx\frac{1}{2}\int n(\textbf{r})\textbf{v}_s^2d\textbf{r}=\frac{m^2}{2}\int \frac{n(\textbf{r})}{\rho^2}d^3\textbf{r}.$$

Using this ansatz (\ref{anzats}) we obtain
the nucleation energy for a vortex with $m=1$ as the function of the vortex core
coordinate $E_\textrm{n}(x_0)$ [see Fig.~\ref{NucleationEnergy}(b)]. We consider the vortex line, shifted along the vector $\textbf{n}$, i.e. into the direction of lowest density of the weak link. As is seen from Fig.~\ref{NucleationEnergy}(b), the nucleation energy has a local minimum at the axis of the trap, when there is no weak link ($U_b=0$).  When $U_b$ grows, the nucleation energy of the vortex within the weak link decays. It is remarkable, that  when a weak link appears the equilibrium position of the vortex core, which corresponds to the local minimum of $E_\textrm{n}(x_0)$, shifts from the center of the ring.

In our time-independent energetic analysis we do not account for the contribution to the vortex energy from the rotation of the weak link.  Obviously, the rotating barrier governs the vortex dynamics by superflows induced in the ring condensate. In the next section we consider dynamics of the phase slips in the framework of the time-dependent dissipative GPE. However, some qualitative predictions can be obtained from analysis of the nucleation energy.

As was pointed out above, the local minima of $E_n(x_0)$ shifts from the axis of the trap in direction of the weak link. Thus, dynamically, one would expect that a negatively charged anti-vortex line approaches the inner surface of the ring condensate and penetrates deeper and deeper into the condensate annulus through the weak link, when the height $U(t)$ of the rotating anti-clockwise barrier grows with time. From the outer edge of the ring the positively charged vortex is expected to penetrate into the weak link. Accounting on the spatial distribution of $E_n(x_0)$ one would expect that the vortex from the external side of the annulus penetrates deeper into the weak link, than the antivortex (having the same energy) from the inner side of the ring does.

Since the two vortices with opposite charges attract each other, a vortex dipole possesses a binding energy. Thus, if the antivortex from the inner side and the vortex from the outer side form a dipole, it gains an additional energy and nonzero linear momentum, while the angular momentum of the vortex dipole is equal to zero. The binding  energy of the moving vortex dipole could be enough to overcome the energetic barrier, weakened by the weak link.  It is reasonable to suggest that the formation of the vortex dipole, which eliminates the negative charge of the inner antivortex, drives the phase slip. These expectations from the energetic stability analysis of the vortex line, are confirmed by our numerical simulations of the time evolution of the condensate discussed in the next section.


\section{Dynamics of the phase slips}

In the experiment described in Ref.~\cite{PhysRevLett.110.025302} the time-averaged barrier is rotated
for 1.5 s at a constant angular velocity $\Omega / 2\pi \le 3\,$Hz, which is well below the angular frequency of
sound propagating around the ring. For slow rotation rates pronounced quantized phase slips are observed at  well-defined critical angular velocities. Here, we numerically simulate the generation of persistent currents via  such phase slips driven by a wide rotating barrier using similar parameters as used  experimental conditions of Ref.~\cite{PhysRevLett.110.025302}.

Furthermore, we consider a broad spectrum of different angular frequencies $\Omega$, including  fast rotation of the barrier, and study the ensuing dynamics of the system. The number of vortices excited in the condensate increases with the rotation speed, and at high speeds we observe turbulent behavior. A toroidal condensate with a rapidly rotating weak link offers intriguing possibilities for the theoretical investigation of quantum turbulence with gain and loss in a quasi-2D geometry~\cite{PhysRevLett.111.235301}.

\subsection{Slowly rotating weak link}

\begin{figure}[ht]
   \includegraphics[width=3.4in]{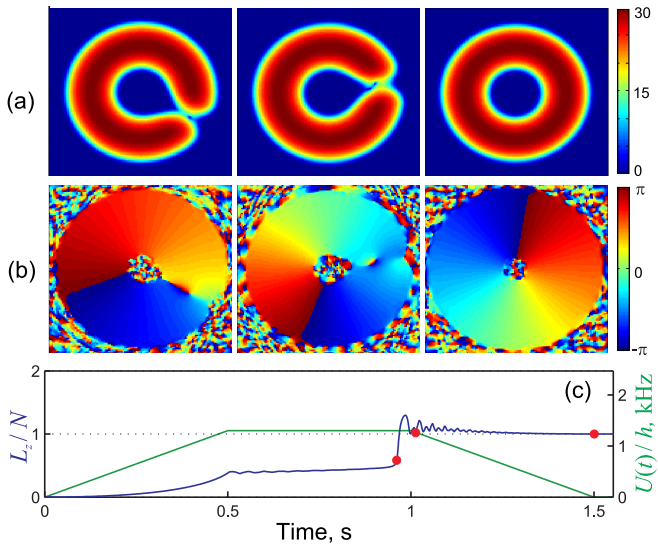}
  \caption{(Color online) Dynamics of the toroidal condensate with a slowly rotating weak link ($\Omega / 2\pi = 2\,$Hz). Snapshots for $t=0.938\,$s, $t=1.013\,$s, and $t=1.5\,$s: (a) $|\Psi(x,y,0)|$;  (b) Arg$\Psi(x,y,0)$. (c) Angular momentum per particle (blue curve) combined with the amplitude of the rotating weak link (green line) as functions of time. Red dots indicate the moments of time corresponding to the snapshots shown in (a) and (b).}
  \label{Dynamics2Hz}
  \end{figure}

Figure~\ref{Dynamics2Hz} illustrates a typical example of a $0\to 1$ phase slip  ($\Omega / 2\pi=  2\,$Hz, $U_b/h=1.3\,$kHz).  A vortex line enters the weak link from the outside and travels towards the central hole. As expected from the predictions of the energetic stability analysis one observes the following: while the vortex traverses the weak link, a negatively-charged anti-vortex line from the central hole and the incoming vortex approach each other and create a vortex-antivortex dipole (see Fig.~\ref{Dynamics2Hz} at $t=0.938\,$s). This coupled pair of vortices circles clockwise inside the central hole
until it reaches the region of the weak link.  The moving dipole usually escapes from the central hole and finally decays, however, during the stirring time the vortex dipole can  several times enter and leave the central hole of the ring through the weak link. If $\Omega$ and $U_b$ are just above some threshold values, vortex dipoles barely have the energy, which is necessary to overcome the potential barrier in the region of the weak link. The vortex is therefore drawn into the central hole of the ring. Thus, when the stirrer begins to ramp down, the low-energetic dipoles remain  trapped and due to dissipation they gradually spiral to the axis of the trap.

 \begin{figure}[ht]
   \includegraphics[width=3.4in]{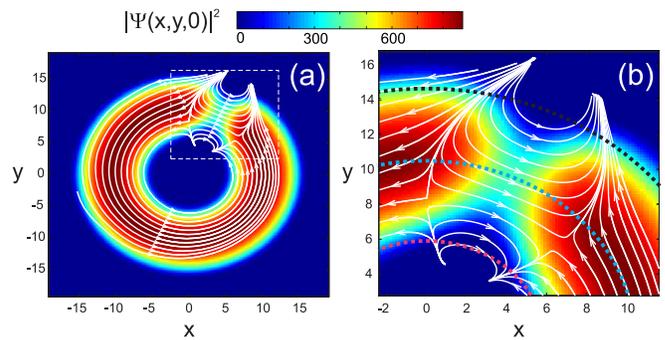}
  \caption{(Color online) (a) Color coded condensate density $|\Psi(x,y,0)|^2$  (with the weak link, which rotates anti-clockwise $\Omega / 2\pi = 3\,$Hz, $U_b=1.3\,$kHz, $t=0.395\,$s) combined with the superflow stream lines. (b) Scaled-up image of the stream lines close to the weak link.}
  \label{StremLines}
  \end{figure}
\begin{figure}[ht]
   \includegraphics[width=3.4in]{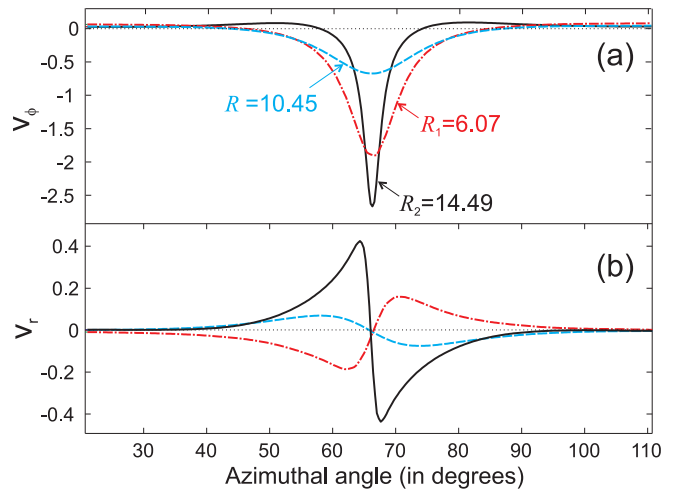}
  \caption{(Color online) The superflow velocities as functions of azimuthal angle $\varphi$: (a) azimuthal velocity $v_\varphi$, (b) radial velocity $v_r$.  The velocities are calculated along the circles (shown in Fig.~\ref{StremLines}~(b) by dotted lines) with radius $R=10.45$ (blue dashed curve), $R_1=6.07$ (red dash-dotted curve), and $R_2=14.49$ (solid black curve).}
  \label{Velocities}
  \end{figure}
If the barrier rotation rate is well above a threshold, several vortices enter into the central hole and form bound pairs with anti-vortices from the central hole. Finally the vortex-antivortex pairs jump out of the condensate. The total charge of the escaping vortex dipole is equal to zero, but each time an external vortex enters the condensate it adds  one unit of topological charge to the global persistent current. In the example labeled  SV2 in Fig.~\ref{LvsOmega} (see the Supplemental Material for an animation) the final state corresponds to a two-charged persistent current.

To gain a better insight into the physical mechanism of the persistent current generation by a wide stirrer it is useful to consider the structure of the superflow just before a phase slip. The velocity field corresponding to $\Omega / 2\pi = 3\,$Hz, $U_b=1.3\,$kHz, $t=0.395\,$s is shown in Figs.~\ref{StremLines} and  \ref{Velocities}.
Note that within the weak link the direction of the azimuthal flow is opposite to the direction of the barrier rotation. It is remarkable, that a forward azimuthal flow appears well in advance of the persistent current formation. This specific feature of the multiply-connected superfluid was observed experimentally and explained qualitatively in Ref.~\cite{arxiv1406_1095} as follows: a low-density wake appearing directly behind the stirrer initiates two counter-propagating superflows: The first atomic flow tends to fill the low-density wake behind the barrier. Because the condensate wave-function must be single valued, the velocity circulation around any closed path is quantized: $M\oint \textbf{v}_s d\textbf{l}=2\pi \hbar m$ and the total velocity circulation vanishes for the state with zero winding number ($m=0$). The second atomic flow occurs in order to cancel the velocity field of the atoms moving through the barrier. 

The azimuthal distribution of the radial $v_r$ and tangential $v_\varphi$ velocities are shown in Fig.~\ref{Velocities}. The superflow velocity $\textbf{v}_s$ has been extracted from the atomic flow $|\psi|^2\textbf{v}_s=-\frac{i}{2}(\psi^*\nabla \psi-\psi\nabla \psi^*)$ at three circles indicated in Fig.~\ref{StremLines} (b) by dots. The blue dotted circle has the radius equal to the $R$ (the toroidal trap radius). Two other circles correspond to the internal (red dots) and external (black dots)  edges of the condensate in the region of the weak link \cite{defineCircles}. It is seen from Fig.~\ref{Velocities} that the superflow moves faster at the external periphery. Since the critical velocity for vortex nucleation is reached first at the outer edge of the condensate, a vortex enters into the weak link rather than an anti-vortex. As we pointed out above,  the azimuthal velocity is positive (forward flow) far from the weak link, but $v_\varphi<0$ inside the weak link (backward flow). Note,  that   at the external edge of the condensate the velocity of the forward flow has the maximums just before and after the weak link. As it is seen from Figs.~\ref{StremLines} and \ref{Velocities}, radial velocity $v_r$ has the different direction at the external and internal edges of the weak link.

It is remarkable that for low rotation rate both a weak link (a barrier which is wider than the condensate annulus) \cite{PhysRevLett.110.025302} and a small (in comparison with the width of the annulus) stirrer used in Ref. \cite{Wright2013} drive the phase slip by excitation of the vortex dipoles. However, the microscopic mechanisms of the phase slip are qualitatively different for these two cases.
As was shown in Ref.~\cite{SmallWL_arxiv14}, a small stirrer at low rotation rate excites vortex-antivortex pairs near the center of the rotating barrier in the bulk of the condensate. Then the pair undergoes a breakdown and the antivortex moves spirally to the external surface of the condensate and finally decays into elementary excitations, while the vortex becomes pinned in the central hole of the annulus adding  one unit to the topological charge of the persistent current.

 It is interesting to compare the mechanism of the persistent current generation  described in the present work with the inverse process of the persistent current decay under the influence of the tunable weak link, which was observed experimentally \cite{PhysRevLett.99.260401,PhysRevLett.106.130401} and investigated theoretically in the framework of the conservative mean field approximation \cite{PhysRevA.80.021601,Piazza2013} and truncated Wigner approximation \cite{TWA}.
In Refs.~\cite{PhysRevA.80.021601,Piazza2013} vortex-antivortex \emph{annihilation} within the weak link caused the phase slip and decay of the persistent current. In our simulations antivortex and vortex are nucleated from the inner and outer side, similar as it was observed in \cite{PhysRevA.80.021601,Piazza2013}, but instead of annihilation with emission of a sound waves, in our dissipative dynamics the moving vortex dipole appears.
The moving dipole plays a key role in the microscopic mechanism of the persistent current generation by slowly rotating weak link. Note that the role of the vortex dipoles in the process of the persistent current generation was pointed out in Ref.~\cite{nature14}, where the energy  of the vortex dipole, which moves inside the rotating weak link, has been analyzed.


 \begin{figure}[ht]
  \includegraphics[width=3.4in]{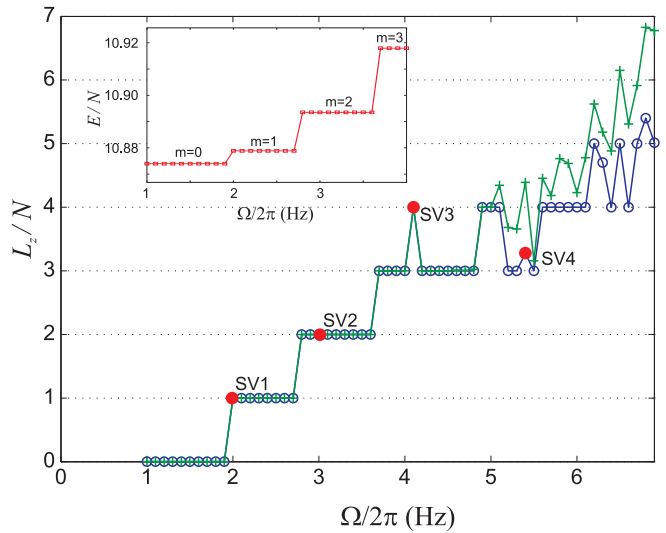}
  \caption{(Color online) Angular momentum per atom as a function of angular velocity for the barrier height $U_b/h=1.3\,$kHz at the moment of time $t=1.5\,$s (green crosses) and for $t=4.5\,$s (open blue circles). Inset presents the condensate energy per atom at $t=1.5\,$s.}
  \label{LvsOmega}
  \end{figure}

In Fig.~\ref{LvsOmega} the angular momentum per particle $L_z/N$ is shown immediately after the end of the stirring
period (green crosses) and after additional 3 seconds (blue open circles).
Here $L_z$ is $z$ component of angular momentum:
$$
\vec L = -
\frac{i}{2}\int \left\{\psi^*[\textbf{r}\times \nabla\psi]-\psi[\textbf{r}\times \nabla\psi^*]\right\}d^3\textbf{r}.
$$
 In qualitative agreement with the experimental observations \cite{PhysRevLett.110.025302} the condensate is in a zero rotation state ($m=L_z/N=0$) for small $\Omega$. When $\Omega$ reaches some critical value, $m$ changes from 0 to 1 and remains constant until the angular velocity reaches the next critical rotation rate, where a transition to $m=2$ occurs. Pronounced quantized phase slips are clearly seen for $\Omega / 2\pi \le 4\,$Hz.

 The inset in Fig.~\ref{LvsOmega} shows the energy per atom immediately after the stirring is stopped ($t=1.5$ s). The well-defined energy spectrum for the states with different winding numbers $m$: $E_m=E_0+\delta E\,m^2$ proves that in these simulations the condensate relaxes to the metastable state during 1.5 s. Here $E_0$ is the energy of the ground state ($m=0$). Note that the  energy cost $\delta E=E_{m=1}-E_0$ for the transition from the ground state to the first excited state (corresponding to the phase slip $0\to 1$) is in excellent agreement with prediction of the energetic stability analysis: $\delta E\approx E_n(x_0=0)=4.9 \cdot 10^{-3}N$, where $E_n(x_0=0)$ is the nucleation energy of the vortex placed at the axis of the trap [see Fig.~\ref{NucleationEnergy} (b)], $N$ is the total number of atoms.

 For higher angular velocities some integer-valued fluctuations appear in $L_z/N$ as the function of $\Omega$.  When $\Omega$ grows further the angular momentum increases on the average, however the specific value of $L_z/N$ at $t=1.5$ s becomes quite unpredictable.

 Under the conditions of the experiment \cite{PhysRevLett.110.025302}, the measured lifetime of the annular vortices is about 3\,s; therefore, an annular vortex formed during the 1.5\,s stirring procedure is clearly seen in TOF imaging. The number of annular vortices, detected in the experiment  after the stirring procedure, increased for higher rotation rate.  Obviously, in our simulations the annular vortices give a non-integer contribution to the winding number $L_z/N$, thus for a rapidly rotating weak link an additional 3\,s holding after the stirring procedure seems to be enough for relaxation of the condensate to a metastable state with $m$-charged persistent current.

 As is seen from Fig.~\ref{LvsOmega} at $t=4.5\,$s relaxation indeed happens, but irregular jumps of the winding number $m$ as a function of $\Omega$ remain even after such a long-term evolution. Moreover, the angular momentum per particle does not reach an integer value for some long-term simulations, which means that the condensate is not yet relaxed to the metastable state. These odd features are the result of the complex dynamics of the annular vortices in the condensate with a rapidly rotating barrier. The point is that more energetic vortices are more long-life, since they penetrates deeper into the bulk of the condensate and interact with another annular vortices. For example, for the simulation labeled SV3 in Fig.~\ref{LvsOmega} (see the Supplemental Material for the animation) one annular vortex moves back and forth from the inner to the outer side of the annulus periodically crossing the weak link. In the simulation labeled SV3 by pure accident the vortex appears to be in the central hole of the toroidal condensate, when the rotating potential barrier was ramping down and the weak link closed. Thus, the winding number in this simulation is  one unit above the expected charge of the persistent current  at this $\Omega$.

 An example for an evolution with anomalously long relaxation time is marked SV4 in Fig.~\ref{LvsOmega} (see the Supplemental Material for an animation). In this simulation  two  positively charged vortices remain deep in the bulk of the condensate, while all other  vortices nucleated by the stirring beam  either decay at the external condensate edge or are pinned at the central hole of the ring. The remaining two vortices exhibit a very complex dynamics governed by the interaction between superflows induced by these vortices. After a long cooperative movement the first of the annular vortices escapes through the outer condensate edge, and only after that the second vortex progressively spirals to the outer periphery of the condensate.

\begin{figure}[ht]
   \includegraphics[width=3.4in]{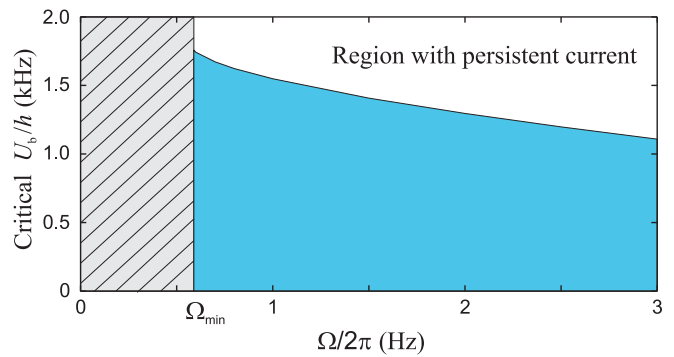}
  \caption{(Color online) Parameter regions ($U_b$,$\Omega$) where no phase slip is observed (shaded regions), $0\to 1$ phase slip is observed (unshaded region).  For frequencies $\Omega<\Omega_\textrm{min}\approx\frac12\Omega_0$ no phase slip is possible for arbitrary barrier intensity.}
  \label{Threshold}
  \end{figure}

We have performed a series of numerical simulations for various combinations of $U_b$ and $\Omega$. To summarize our finding for slow rotation of the stirrer, we present  in Fig.~\ref{Threshold} the threshold for $0\to 1$ phase slip in the plane of the weak link parameters: barrier height $U_b/h$ and $\Omega/(2\pi)$. The critical barrier height $U_b/h$ for each $\Omega$ shown in Fig.~\ref{Threshold} was obtained with accuracy not worth than 5\,Hz. In qualitative agreement with the experiment, the threshold barrier height increases when the angular velocity decreases. Particularly striking is that below a well-defined angular velocity $\Omega_\textrm{min}/2\pi \approx  0.58\,$Hz the phase slip does not occur at any barrier height $U_b$. The value of minimal angular velocity appears to be close to predictions of Ref.~\cite{nature14}: $\Omega_\textrm{min}\approx \Omega_0/2$, where $\Omega_0=\hbar/(M R^2)= 2 \pi \times  1.13\,$Hz is the rotational quantum. It turns out that formation of the $m=1$ persistent currents becomes energetically unfavorable for $\Omega<\Omega_\textrm{min}$.

\subsection{Rapidly rotating weak link}
For higher angular velocities of the rotating barrier the mechanism of the phase slips is quite different.
To get a better insight into the mechanism of the excitation of the vortices in the 3D toroidal condensate, it is useful to consider 2D excitations of the annular condensate.
As is well known, in the annular geometry the radial degrees of freedom lead to specific features of the excitations, such as inner and outer edge surface modes investigated previously for 2D case in Refs.~\cite{PhysRevA.86.011604, PhysRevA.86.011602}. The critical frequency can be obtained by an analysis of small azimuthal perturbations around the stationary state. Alternatively, it may be obtained  from a simple surface model \cite{PhysRevA.86.011602} $\Omega_c/\omega_r=\sqrt{2}\mu^{1/6}/(R+\Delta R/2)$. Both approximations give practically the same result: $\Omega_c/2\pi \approx 17.14\,$Hz for the parameters of our model.

\begin{figure}[ht]
   \includegraphics[width=3.4in]{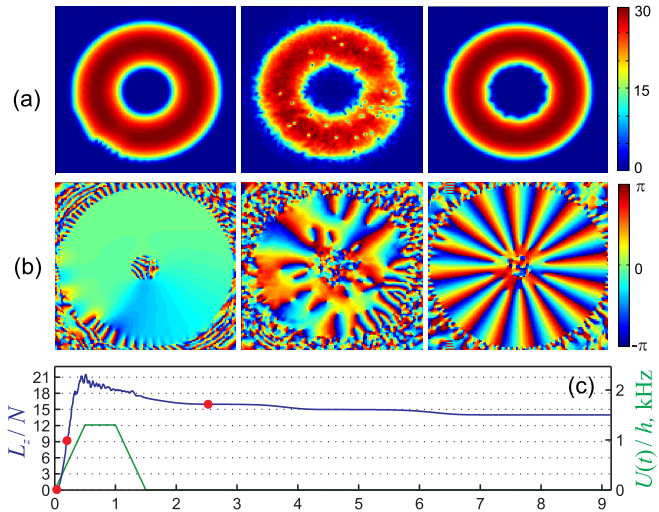}
  \caption{(Color online) Dynamics of the toroidal condensate with rapidly rotating weak link ($\Omega / 2\pi = 20\,$Hz). Snapshots for $t=0.034\,$s, $t=0.2\,$s, and $t=2.52\,$s: (a) $|\Psi(x,y,0)|$;  (b) Arg$\Psi(x,y,0)$. (c) Angular momentum per particle (blue curve) combined with the amplitude of the rotating weak link (green line) as functions of time. Red dots indicate the moments of time, shown in the snapshots (a) and (b).}
  \label{Dynamics20Hz}
  \end{figure}

 In our 3D dynamical simulations we observed that if the angular velocity is above $\Omega_c$, surface waves indeed develop at the external surface of the annulus in the region of the weak link (see Fig. \ref{Dynamics20Hz}). The corresponding surface wave breaks and vortices subsequently enter the bulk of the condensate. But it turns out, that the ripples at the outer edge of the weak link develop also below $\Omega_c$ and we have not observed a well-defined threshold angular frequency for transition to the regime with external surface modes.

The number of annular vortices and the topological charge of the supercurrent rapidly increase with $\Omega$, so that the dynamics of the vortices becomes quite irregular.  The complex dynamics of the vortices is governed by  condensate inhomogeneities, dissipation effects, and by the interplay between
condensate flows corresponding to other vortices. An example of the time evolution of the toroidal condensate for very large $\Omega$ is shown in Fig.~\ref{Dynamics20Hz}. 

It is remarkable, that highly-charged persistent currents, which are generated by the stirring procedure can be unstable with respect to serial \textit{emission} of vortex lines. During such an emission process a vortex line spirals out from the central hole of the toroid through the bulk of the condensate without weak link. It is interesting to observe that the series of the vortex emissions begins usually for a long-term evolution, when the bulk of the condensate is already clear of vortices (see Fig.~\ref{Dynamics20Hz}).

Note that an internal surface wave develops before the emitted vortex penetrates into the bulk of the condensate from the central hole.
The persistent current, which is prepared by stirring with rotating weak link, is not exactly in a metastable state, which corresponds to $m$-charge vortex line at the trap axis. Obviously, it has an additional energy stored in the bound off-set vortices, which are trapped in the central hole. This energy could be transferred first to the collective inner surface modes  and then to the emitted vortex, which induces the phase slip.
Indeed, as is seen from Fig.~\ref{Dynamics20Hz} for $t=2.52\,$s,  a highly-charged persistent current can be unstable with respect to excitations of the internal surface waves, and  a spiraling annular vortex appears when the internal surface wave significantly grows and then breaks.

\section{Discussion and conclusions}

 Our theoretical results are in qualitative agreement with the experimental findings: the threshold angular velocity decreases, when the barrier height increases. However, comparison of the two experimental series for the fixed barrier height $U_b$ with Fig.~\ref{Threshold} shows that theoretically predicted $\Omega$ is significantly higher than experimentally measured angular velocity for the phase slip $0\to 1$. A considerable discrepancy between the experiment and theoretical results obtained in the framework of 3D dissipative GPE was also reported in Ref.~\cite{nature14} for the hysteresis effect in toroidal condensate. It was assumed in Ref.~\cite{nature14} that more sophisticated model, which includes a variation of the chemical potential with time could resolve this description. As we described above, our model properly describes variation of the chemical potential with time and reproduces a typical experimental decay rate of the number of atoms with time. Nevertheless, the calculated threshold values of $U_b$ and $\Omega$ do not meet a quantative correspondence with experimental measurements of Ref.~\cite{PhysRevLett.110.025302}.

More theoretical and experimental work is needed to reveal the physical mechanisms, which are not taken into account in a simple dissipative mean-field theory.
In accordance with experimental conditions, we assume that the temperature of the condensate is well below the condensation temperature $T_c$, thus vortices are only weakly affected by thermal noise. Numerous experimental and theoretical investigations demonstrated, that the dissipative GPE correctly describes the averaged dynamics of the vortices in single-connected and toroidal condensate when $T\ll T_c$.
However, the stochastic thermal fluctuations could significantly increase the probability of the phase slips, when the weak link suppresses the potential barrier for vortices in the ring-shaped condensate. The influence of the thermal fluctuations on the excitation and dynamics of the vortices in a toroidal trap with a rotating weak link merits a separate study. The reported in experiments shot-to-shot atom number (and, probably, temperature) fluctuations as well as considerable impurities in the condensate density distribution also need a detailed analysis.

We believe that the mechanism of the phase slips, described in the present work, can be similarly observed in toroidal BECs in the hysteresis regime \cite{nature14} and in multijunction atomic circuits \cite{PhysRevLett.113.045305}.




\begin{thebibliography}{30}
\expandafter\ifx\csname natexlab\endcsname\relax\def\natexlab#1{#1}\fi
\expandafter\ifx\csname bibnamefont\endcsname\relax
  \def\bibnamefont#1{#1}\fi
\expandafter\ifx\csname bibfnamefont\endcsname\relax
  \def\bibfnamefont#1{#1}\fi
\expandafter\ifx\csname citenamefont\endcsname\relax
  \def\citenamefont#1{#1}\fi
\expandafter\ifx\csname url\endcsname\relax
  \def\url#1{\texttt{#1}}\fi
\expandafter\ifx\csname urlprefix\endcsname\relax\def\urlprefix{URL }\fi
\providecommand{\bibinfo}[2]{#2}
\providecommand{\eprint}[2][]{\url{#2}}

\bibitem[{\citenamefont{{Ryu} et~al.}(2007)\citenamefont{{Ryu}, {Andersen},
  {Clad{\'e}}, {Natarajan}, {Helmerson}, and
  {Phillips}}}]{PhysRevLett.99.260401}
\bibinfo{author}{\bibfnamefont{C.}~\bibnamefont{{Ryu}}},
  \bibinfo{author}{\bibfnamefont{M.~F.} \bibnamefont{{Andersen}}},
  \bibinfo{author}{\bibfnamefont{P.}~\bibnamefont{{Clad{\'e}}}},
  \bibinfo{author}{\bibfnamefont{V.}~\bibnamefont{{Natarajan}}},
  \bibinfo{author}{\bibfnamefont{K.}~\bibnamefont{{Helmerson}}},
  \bibnamefont{and} \bibinfo{author}{\bibfnamefont{W.~D.}
  \bibnamefont{{Phillips}}}, \bibinfo{journal}{\prl}
  \textbf{\bibinfo{volume}{99}}, \bibinfo{eid}{260401} (\bibinfo{year}{2007}).

\bibitem[{\citenamefont{{Ramanathan} et~al.}(2011)\citenamefont{{Ramanathan},
  {Wright}, {Muniz}, {Zelan}, {Hill}, {Lobb}, {Helmerson}, {Phillips}, and
  {Campbell}}}]{PhysRevLett.106.130401}
\bibinfo{author}{\bibfnamefont{A.}~\bibnamefont{{Ramanathan}}},
  \bibinfo{author}{\bibfnamefont{K.~C.} \bibnamefont{{Wright}}},
  \bibinfo{author}{\bibfnamefont{S.~R.} \bibnamefont{{Muniz}}},
  \bibinfo{author}{\bibfnamefont{M.}~\bibnamefont{{Zelan}}},
  \bibinfo{author}{\bibfnamefont{W.~T.} \bibnamefont{{Hill}},
  \bibfnamefont{III}}, \bibinfo{author}{\bibfnamefont{C.~J.}
  \bibnamefont{{Lobb}}},
  \bibinfo{author}{\bibfnamefont{K.}~\bibnamefont{{Helmerson}}},
  \bibinfo{author}{\bibfnamefont{W.~D.} \bibnamefont{{Phillips}}},
  \bibnamefont{and} \bibinfo{author}{\bibfnamefont{G.~K.}
  \bibnamefont{{Campbell}}}, \bibinfo{journal}{\prl}
  \textbf{\bibinfo{volume}{106}}, \bibinfo{eid}{130401} (\bibinfo{year}{2011}).

\bibitem[{\citenamefont{Wright et~al.}(2013)\citenamefont{Wright, Blakestad,
  Lobb, Phillips, and Campbell}}]{PhysRevLett.110.025302}
\bibinfo{author}{\bibfnamefont{K.~C.} \bibnamefont{Wright}},
  \bibinfo{author}{\bibfnamefont{R.~B.} \bibnamefont{Blakestad}},
  \bibinfo{author}{\bibfnamefont{C.~J.} \bibnamefont{Lobb}},
  \bibinfo{author}{\bibfnamefont{W.~D.} \bibnamefont{Phillips}},
  \bibnamefont{and} \bibinfo{author}{\bibfnamefont{G.~K.}
  \bibnamefont{Campbell}}, \bibinfo{journal}{Phys. Rev. Lett.}
  \textbf{\bibinfo{volume}{110}}, \bibinfo{pages}{025302}
  (\bibinfo{year}{2013}).

\bibitem[{\citenamefont{Jendrzejewski et~al.}(2014)\citenamefont{Jendrzejewski,
  Eckel, Murray, Lanier, Edwards, Lobb, and Campbell}}]{PhysRevLett.113.045305}
\bibinfo{author}{\bibfnamefont{F.}~\bibnamefont{Jendrzejewski}},
  \bibinfo{author}{\bibfnamefont{S.}~\bibnamefont{Eckel}},
  \bibinfo{author}{\bibfnamefont{N.}~\bibnamefont{Murray}},
  \bibinfo{author}{\bibfnamefont{C.}~\bibnamefont{Lanier}},
  \bibinfo{author}{\bibfnamefont{M.}~\bibnamefont{Edwards}},
  \bibinfo{author}{\bibfnamefont{C.~J.} \bibnamefont{Lobb}}, \bibnamefont{and}
  \bibinfo{author}{\bibfnamefont{G.~K.} \bibnamefont{Campbell}},
  \bibinfo{journal}{Phys. Rev. Lett.} \textbf{\bibinfo{volume}{113}},
  \bibinfo{pages}{045305} (\bibinfo{year}{2014}).

\bibitem[{\citenamefont{Ryu et~al.}(2013)\citenamefont{Ryu, Blackburn, Blinova,
  and Boshier}}]{PhysRevLett.111.205301}
\bibinfo{author}{\bibfnamefont{C.}~\bibnamefont{Ryu}},
  \bibinfo{author}{\bibfnamefont{P.~W.} \bibnamefont{Blackburn}},
  \bibinfo{author}{\bibfnamefont{A.~A.} \bibnamefont{Blinova}},
  \bibnamefont{and} \bibinfo{author}{\bibfnamefont{M.~G.}
  \bibnamefont{Boshier}}, \bibinfo{journal}{Phys. Rev. Lett.}
  \textbf{\bibinfo{volume}{111}}, \bibinfo{pages}{205301}
  (\bibinfo{year}{2013}).

\bibitem[{\citenamefont{Kasamatsu et~al.}(2002)\citenamefont{Kasamatsu,
  Tsubota, and Ueda}}]{PhysRevA.66.053606}
\bibinfo{author}{\bibfnamefont{K.}~\bibnamefont{Kasamatsu}},
  \bibinfo{author}{\bibfnamefont{M.}~\bibnamefont{Tsubota}}, \bibnamefont{and}
  \bibinfo{author}{\bibfnamefont{M.}~\bibnamefont{Ueda}},
  \bibinfo{journal}{Phys. Rev. A} \textbf{\bibinfo{volume}{66}},
  \bibinfo{pages}{053606} (\bibinfo{year}{2002}).

\bibitem[{\citenamefont{{Benakli} et~al.}(1999)\citenamefont{{Benakli},
  {Raghavan}, {Smerzi}, {Fantoni}, and {Shenoy}}}]{BenakliEuL99}
\bibinfo{author}{\bibfnamefont{M.}~\bibnamefont{{Benakli}}},
  \bibinfo{author}{\bibfnamefont{S.}~\bibnamefont{{Raghavan}}},
  \bibinfo{author}{\bibfnamefont{A.}~\bibnamefont{{Smerzi}}},
  \bibinfo{author}{\bibfnamefont{S.}~\bibnamefont{{Fantoni}}},
  \bibnamefont{and} \bibinfo{author}{\bibfnamefont{S.~R.}
  \bibnamefont{{Shenoy}}}, \bibinfo{journal}{EPL (Europhysics Letters)}
  \textbf{\bibinfo{volume}{46}}, \bibinfo{pages}{275} (\bibinfo{year}{1999}).

\bibitem[{\citenamefont{{Brand} and {Reinhardt}}(2001)}]{Brand01}
\bibinfo{author}{\bibfnamefont{J.}~\bibnamefont{{Brand}}} \bibnamefont{and}
  \bibinfo{author}{\bibfnamefont{W.~P.} \bibnamefont{{Reinhardt}}},
  \bibinfo{journal}{Journal of Physics B Atomic Molecular Physics}
  \textbf{\bibinfo{volume}{34}}, \bibinfo{pages}{L113} (\bibinfo{year}{2001}).

\bibitem[{\citenamefont{{Das} et~al.}(2012)\citenamefont{{Das}, {Sabbatini},
  and {Zurek}}}]{Das2012}
\bibinfo{author}{\bibfnamefont{A.}~\bibnamefont{{Das}}},
  \bibinfo{author}{\bibfnamefont{J.}~\bibnamefont{{Sabbatini}}},
  \bibnamefont{and} \bibinfo{author}{\bibfnamefont{W.~H.}
  \bibnamefont{{Zurek}}}, \bibinfo{journal}{Scientific Reports}
  \textbf{\bibinfo{volume}{2}}, \bibinfo{eid}{352} (\bibinfo{year}{2012}),
  \eprint{1102.5474}.

\bibitem[{\citenamefont{Martikainen et~al.}(2001)\citenamefont{Martikainen,
  Suominen, Santos, Schulte, and Sanpera}}]{PhysRevA.64.063602}
\bibinfo{author}{\bibfnamefont{J.-P.} \bibnamefont{Martikainen}},
  \bibinfo{author}{\bibfnamefont{K.-A.} \bibnamefont{Suominen}},
  \bibinfo{author}{\bibfnamefont{L.}~\bibnamefont{Santos}},
  \bibinfo{author}{\bibfnamefont{T.}~\bibnamefont{Schulte}}, \bibnamefont{and}
  \bibinfo{author}{\bibfnamefont{A.}~\bibnamefont{Sanpera}},
  \bibinfo{journal}{Phys. Rev. A} \textbf{\bibinfo{volume}{64}},
  \bibinfo{pages}{063602} (\bibinfo{year}{2001}).

\bibitem[{\citenamefont{Modugno et~al.}(2006)\citenamefont{Modugno, Tozzo, and
  Dalfovo}}]{PhysRevA.74.061601}
\bibinfo{author}{\bibfnamefont{M.}~\bibnamefont{Modugno}},
  \bibinfo{author}{\bibfnamefont{C.}~\bibnamefont{Tozzo}}, \bibnamefont{and}
  \bibinfo{author}{\bibfnamefont{F.}~\bibnamefont{Dalfovo}},
  \bibinfo{journal}{Phys. Rev. A} \textbf{\bibinfo{volume}{74}},
  \bibinfo{pages}{061601} (\bibinfo{year}{2006}).

\bibitem[{\citenamefont{Beattie et~al.}(2013)\citenamefont{Beattie, Moulder,
  Fletcher, and Hadzibabic}}]{PhysRevLett.110.025301}
\bibinfo{author}{\bibfnamefont{S.}~\bibnamefont{Beattie}},
  \bibinfo{author}{\bibfnamefont{S.}~\bibnamefont{Moulder}},
  \bibinfo{author}{\bibfnamefont{R.~J.} \bibnamefont{Fletcher}},
  \bibnamefont{and}
  \bibinfo{author}{\bibfnamefont{Z.}~\bibnamefont{Hadzibabic}},
  \bibinfo{journal}{Phys. Rev. Lett.} \textbf{\bibinfo{volume}{110}},
  \bibinfo{pages}{025301} (\bibinfo{year}{2013}).

\bibitem[{\citenamefont{Yakimenko
  et~al.}(2013{\natexlab{a}})\citenamefont{Yakimenko, Isaieva, Vilchinskii, and
  Weyrauch}}]{PRA2013R}
\bibinfo{author}{\bibfnamefont{A.~I.} \bibnamefont{Yakimenko}},
  \bibinfo{author}{\bibfnamefont{K.~O.} \bibnamefont{Isaieva}},
  \bibinfo{author}{\bibfnamefont{S.~I.} \bibnamefont{Vilchinskii}},
  \bibnamefont{and} \bibinfo{author}{\bibfnamefont{M.}~\bibnamefont{Weyrauch}},
  \bibinfo{journal}{Phys. Rev. A} \textbf{\bibinfo{volume}{88}},
  \bibinfo{pages}{051602} (\bibinfo{year}{2013}{\natexlab{a}}).

\bibitem[{\citenamefont{Mason and Berloff}(2009)}]{Berloff09}
\bibinfo{author}{\bibfnamefont{P.}~\bibnamefont{Mason}} \bibnamefont{and}
  \bibinfo{author}{\bibfnamefont{N.~G.} \bibnamefont{Berloff}},
  \bibinfo{journal}{Phys. Rev. A} \textbf{\bibinfo{volume}{79}},
  \bibinfo{pages}{043620} (\bibinfo{year}{2009}).

\bibitem[{\citenamefont{Moulder et~al.}(2012)\citenamefont{Moulder, Beattie,
  Smith, Tammuz, and Hadzibabic}}]{PhysRevA.86.013629}
\bibinfo{author}{\bibfnamefont{S.}~\bibnamefont{Moulder}},
  \bibinfo{author}{\bibfnamefont{S.}~\bibnamefont{Beattie}},
  \bibinfo{author}{\bibfnamefont{R.~P.} \bibnamefont{Smith}},
  \bibinfo{author}{\bibfnamefont{N.}~\bibnamefont{Tammuz}}, \bibnamefont{and}
  \bibinfo{author}{\bibfnamefont{Z.}~\bibnamefont{Hadzibabic}},
  \bibinfo{journal}{Phys. Rev. A} \textbf{\bibinfo{volume}{86}},
  \bibinfo{pages}{013629} (\bibinfo{year}{2012}).

\bibitem[{\citenamefont{Piazza et~al.}(2009)\citenamefont{Piazza, Collins, and
  Smerzi}}]{PhysRevA.80.021601}
\bibinfo{author}{\bibfnamefont{F.}~\bibnamefont{Piazza}},
  \bibinfo{author}{\bibfnamefont{L.~A.} \bibnamefont{Collins}},
  \bibnamefont{and} \bibinfo{author}{\bibfnamefont{A.}~\bibnamefont{Smerzi}},
  \bibinfo{journal}{Phys. Rev. A} \textbf{\bibinfo{volume}{80}},
  \bibinfo{pages}{021601} (\bibinfo{year}{2009}).

\bibitem[{\citenamefont{{Piazza} et~al.}(2013)\citenamefont{{Piazza},
  {Collins}, and {Smerzi}}}]{Piazza2013}
\bibinfo{author}{\bibfnamefont{F.}~\bibnamefont{{Piazza}}},
  \bibinfo{author}{\bibfnamefont{L.~A.} \bibnamefont{{Collins}}},
  \bibnamefont{and} \bibinfo{author}{\bibfnamefont{A.}~\bibnamefont{{Smerzi}}},
  \bibinfo{journal}{Journal of Physics B Atomic Molecular Physics}
  \textbf{\bibinfo{volume}{46}}, \bibinfo{eid}{095302} (\bibinfo{year}{2013}).

\bibitem[{\citenamefont{{Wright} et~al.}(2013)\citenamefont{{Wright},
  {Blakestad}, {Lobb}, {Phillips}, and {Campbell}}}]{Wright2013}
\bibinfo{author}{\bibfnamefont{K.~C.} \bibnamefont{{Wright}}},
  \bibinfo{author}{\bibfnamefont{R.~B.} \bibnamefont{{Blakestad}}},
  \bibinfo{author}{\bibfnamefont{C.~J.} \bibnamefont{{Lobb}}},
  \bibinfo{author}{\bibfnamefont{W.~D.} \bibnamefont{{Phillips}}},
  \bibnamefont{and} \bibinfo{author}{\bibfnamefont{G.~K.}
  \bibnamefont{{Campbell}}}, \bibinfo{journal}{\pra}
  \textbf{\bibinfo{volume}{88}}, \bibinfo{eid}{063633} (\bibinfo{year}{2013}).

\bibitem[{\citenamefont{Pitaevskii}(1959)}]{Pitaevskii59}
\bibinfo{author}{\bibfnamefont{L.}~\bibnamefont{Pitaevskii}},
  \bibinfo{journal}{Sov. Phys. JETP} \textbf{\bibinfo{volume}{8}},
  \bibinfo{pages}{88} (\bibinfo{year}{1959}).

\bibitem[{\citenamefont{Choi et~al.}(1998)\citenamefont{Choi, Morgan, and
  Burnett}}]{Choi98}
\bibinfo{author}{\bibfnamefont{S.}~\bibnamefont{Choi}},
  \bibinfo{author}{\bibfnamefont{S.~A.} \bibnamefont{Morgan}},
  \bibnamefont{and} \bibinfo{author}{\bibfnamefont{K.}~\bibnamefont{Burnett}},
  \bibinfo{journal}{Phys. Rev. A} \textbf{\bibinfo{volume}{57}},
  \bibinfo{pages}{4057} (\bibinfo{year}{1998}).

\bibitem[{\citenamefont{Jackson et~al.}(1999)\citenamefont{Jackson, McCann, and
  Adams}}]{Jackson99}
\bibinfo{author}{\bibfnamefont{B.}~\bibnamefont{Jackson}},
  \bibinfo{author}{\bibfnamefont{J.~F.} \bibnamefont{McCann}},
  \bibnamefont{and} \bibinfo{author}{\bibfnamefont{C.~S.} \bibnamefont{Adams}},
  \bibinfo{journal}{Phys. Rev. A} \textbf{\bibinfo{volume}{61}},
  \bibinfo{pages}{013604} (\bibinfo{year}{1999}).

\bibitem[{\citenamefont{Yakimenko
  et~al.}(2013{\natexlab{b}})\citenamefont{Yakimenko, Bidasyuk, Prikhodko,
  Vilchinskii, Ostrovskaya, and Kivshar}}]{PRA13}
\bibinfo{author}{\bibfnamefont{A.~I.} \bibnamefont{Yakimenko}},
  \bibinfo{author}{\bibfnamefont{Y.~M.} \bibnamefont{Bidasyuk}},
  \bibinfo{author}{\bibfnamefont{O.~O.} \bibnamefont{Prikhodko}},
  \bibinfo{author}{\bibfnamefont{S.~I.} \bibnamefont{Vilchinskii}},
  \bibinfo{author}{\bibfnamefont{E.~A.} \bibnamefont{Ostrovskaya}},
  \bibnamefont{and} \bibinfo{author}{\bibfnamefont{Y.~S.}
  \bibnamefont{Kivshar}}, \bibinfo{journal}{\pra}
  \textbf{\bibinfo{volume}{88}}, \bibinfo{pages}{043637}
  (\bibinfo{year}{2013}{\natexlab{b}}).

\bibitem[{\citenamefont{Neely et~al.}(2013)\citenamefont{Neely, Bradley,
  Samson, Rooney, Wright, Law, Carretero-Gonz\'alez, Kevrekidis, Davis, and
  Anderson}}]{PhysRevLett.111.235301}
\bibinfo{author}{\bibfnamefont{T.~W.} \bibnamefont{Neely}},
  \bibinfo{author}{\bibfnamefont{A.~S.} \bibnamefont{Bradley}},
  \bibinfo{author}{\bibfnamefont{E.~C.} \bibnamefont{Samson}},
  \bibinfo{author}{\bibfnamefont{S.~J.} \bibnamefont{Rooney}},
  \bibinfo{author}{\bibfnamefont{E.~M.} \bibnamefont{Wright}},
  \bibinfo{author}{\bibfnamefont{K.~J.~H.} \bibnamefont{Law}},
  \bibinfo{author}{\bibfnamefont{R.}~\bibnamefont{Carretero-Gonz\'alez}},
  \bibinfo{author}{\bibfnamefont{P.~G.} \bibnamefont{Kevrekidis}},
  \bibinfo{author}{\bibfnamefont{M.~J.} \bibnamefont{Davis}}, \bibnamefont{and}
  \bibinfo{author}{\bibfnamefont{B.~P.} \bibnamefont{Anderson}},
  \bibinfo{journal}{Phys. Rev. Lett.} \textbf{\bibinfo{volume}{111}},
  \bibinfo{pages}{235301} (\bibinfo{year}{2013}).

\bibitem[{\citenamefont{{Eckel}
  et~al.}(2014{\natexlab{a}})\citenamefont{{Eckel}, {Jendrzejewski}, {Kumar},
  {Lobb}, and {Campbell}}}]{arxiv1406_1095}
\bibinfo{author}{\bibfnamefont{S.}~\bibnamefont{{Eckel}}},
  \bibinfo{author}{\bibfnamefont{F.}~\bibnamefont{{Jendrzejewski}}},
  \bibinfo{author}{\bibfnamefont{A.}~\bibnamefont{{Kumar}}},
  \bibinfo{author}{\bibfnamefont{C.~J.} \bibnamefont{{Lobb}}},
  \bibnamefont{and} \bibinfo{author}{\bibfnamefont{G.~K.}
  \bibnamefont{{Campbell}}}, \bibinfo{journal}{ArXiv e-prints}
  (\bibinfo{year}{2014}{\natexlab{a}}), \eprint{1406.1095}.

\bibitem[{def()}]{defineCircles}
\bibinfo{note}{If $n_b$ is the maximum condensate density along the vector
  $\textbf{n}$ [see Eq. (\ref{straigthBeam})], then external and internal
  circles correspond to the radiuses, where along the direction of $\textbf{n}$
  the density density is only $10^{-3}$ of $n_b$.}

\bibitem[{\citenamefont{{Yakimenko} et~al.}(2014)\citenamefont{{Yakimenko},
  {Isaieva}, {Vilchinskii}, and {Ostrovskaya}}}]{SmallWL_arxiv14}
\bibinfo{author}{\bibfnamefont{A.~I.} \bibnamefont{{Yakimenko}}},
  \bibinfo{author}{\bibfnamefont{K.~O.} \bibnamefont{{Isaieva}}},
  \bibinfo{author}{\bibfnamefont{S.~I.} \bibnamefont{{Vilchinskii}}},
  \bibnamefont{and} \bibinfo{author}{\bibfnamefont{E.~A.}
  \bibnamefont{{Ostrovskaya}}}, \bibinfo{journal}{ArXiv e-prints}
  (\bibinfo{year}{2014}), \eprint{1408.3293}.

\bibitem[{\citenamefont{{Mathey} et~al.}(2014)\citenamefont{{Mathey}, {Clark},
  and {Mathey}}}]{TWA}
\bibinfo{author}{\bibfnamefont{A.~C.} \bibnamefont{{Mathey}}},
  \bibinfo{author}{\bibfnamefont{C.~W.} \bibnamefont{{Clark}}},
  \bibnamefont{and} \bibinfo{author}{\bibfnamefont{L.}~\bibnamefont{{Mathey}}},
  \bibinfo{journal}{\pra} \textbf{\bibinfo{volume}{90}}, \bibinfo{eid}{023604}
  (\bibinfo{year}{2014}).

\bibitem[{\citenamefont{{Eckel}
  et~al.}(2014{\natexlab{b}})\citenamefont{{Eckel}, {Lee}, {Jendrzejewski},
  {Murray}, {Clark}, {Lobb}, {Phillips}, {Edwards}, and {Campbell}}}]{nature14}
\bibinfo{author}{\bibfnamefont{S.}~\bibnamefont{{Eckel}}},
  \bibinfo{author}{\bibfnamefont{J.~G.} \bibnamefont{{Lee}}},
  \bibinfo{author}{\bibfnamefont{F.}~\bibnamefont{{Jendrzejewski}}},
  \bibinfo{author}{\bibfnamefont{N.}~\bibnamefont{{Murray}}},
  \bibinfo{author}{\bibfnamefont{C.~W.} \bibnamefont{{Clark}}},
  \bibinfo{author}{\bibfnamefont{C.~J.} \bibnamefont{{Lobb}}},
  \bibinfo{author}{\bibfnamefont{W.~D.} \bibnamefont{{Phillips}}},
  \bibinfo{author}{\bibfnamefont{M.}~\bibnamefont{{Edwards}}},
  \bibnamefont{and} \bibinfo{author}{\bibfnamefont{G.~K.}
  \bibnamefont{{Campbell}}}, \bibinfo{journal}{Nature}
  \textbf{\bibinfo{volume}{506}}, \bibinfo{pages}{200}
  (\bibinfo{year}{2014}{\natexlab{b}}).

\bibitem[{\citenamefont{Woo and Son}(2012)}]{PhysRevA.86.011604}
\bibinfo{author}{\bibfnamefont{S.~J.} \bibnamefont{Woo}} \bibnamefont{and}
  \bibinfo{author}{\bibfnamefont{Y.-W.} \bibnamefont{Son}},
  \bibinfo{journal}{Phys. Rev. A} \textbf{\bibinfo{volume}{86}},
  \bibinfo{pages}{011604} (\bibinfo{year}{2012}).

\bibitem[{\citenamefont{Dubessy et~al.}(2012)\citenamefont{Dubessy, Liennard,
  Pedri, and Perrin}}]{PhysRevA.86.011602}
\bibinfo{author}{\bibfnamefont{R.}~\bibnamefont{Dubessy}},
  \bibinfo{author}{\bibfnamefont{T.}~\bibnamefont{Liennard}},
  \bibinfo{author}{\bibfnamefont{P.}~\bibnamefont{Pedri}}, \bibnamefont{and}
  \bibinfo{author}{\bibfnamefont{H.}~\bibnamefont{Perrin}},
  \bibinfo{journal}{Phys. Rev. A} \textbf{\bibinfo{volume}{86}},
  \bibinfo{pages}{011602} (\bibinfo{year}{2012}).

\end{thebibliography}

\end{document}